\documentclass[letterpaper,aastex]{emulateapj}
\usepackage{apjfonts}

\def\apj{ ApJ}
\def\aap{ A\&A}
\def\mnras{MNRAS}

\def\araa{ARAA}

\def\apjs{ApJ Supp}

\shorttitle{LMXBs with pre-main sequence companions}
\shortauthors{Ivanova}

\begin{document}



\title{Low mass X-ray binaries with pre-main sequence companions}
\author{N.\ Ivanova$^1$}
\altaffiltext{1}{Canadian Institute for Theoretical Astrophysics, University of Toronto, 60 St. George, Toronto, ON M5S 3H8, Canada}

\begin{abstract}{
In this Letter we examine the idea that a subset of short-period black-hole low-mass 
X-ray binaries  could be powered by the mass transfer from pre-main sequence donors. 
As the star contracts towards the main sequence, the strong magnetic fields operate the magnetic braking which
dissipates the orbital angular momentum, driving the binary to contact.
We show that the periods and apparent donor spectral classes of the X-ray binaries with a pre-main sequence donor
agree better with the available observations of black hole X-ray binaries than those 
of binaries with a main-sequence donor.
This mechanism also explains, without the need for additional hypotheses, the roughly primordial abundance of Li 
detected in donor companions of black hole X-ray candidates in our Galaxy.
}
\end{abstract}


\section{Introduction}

Low mass  X-ray binaries (LMXBs) are mass-transferring (MT) binaries that consist of low mass donor
and a compact accretor, such as a black hole  (BH) or a neutron star. 
Out of 17 detected black-hole X-ray binaries, 11 have relatively short periods, $\la 15$~h 
and 9 of them have other properties measured \citep[for reviews, see][]{Casares05_bhm,Remil06_bh}.
The effective temperatures of donor companions in those binaries suggest that they are
low-mass stars, and their spectral classes indicate that they are not giants. 
In addition, these short-period LMXBs are transient and in those of them
where the lithium abundance was measured (GS 2000+25, A0620-00 and Nova Mus 91), observations show the excess above the solar 
value by a factor of about 20-200 \citep[][ and references
within]{Macc05_li}.

There are several problems with creating a complete picture of the formation and evolution of a LMXB.
The formation scenario for such binary needs to satisfy, at the very least, the following obvious conditions:  
(i) the binary has to survive both the supernova explosion and the pre-supernova binary evolution; and 
(ii) the post-explosion binary has to be close enough to start the mass transfer.
If one requires the MT to start during the main sequence (MS) evolution of the secondary,
then the latter condition limits the maximal post-explosion binary period to about a day (depending
on the black hole mass): in binaries with a bigger period, the magnetic braking will not provide
the necessary orbital shrinkage.

In the standard scenario for the formation of a compact accreting 
BH binary, a common envelope occurrence during the pre-supernova evolution is required to effect the orbital shrinkage.
The balance between the binding energy of the envelope of a massive super-giant and the released orbital energy 
suggests that a low mass donor may have difficulties to eject the entire envelope \citep[see][Appendix]{Justham06_lmxb}, 
unless a significant fraction of the envelope had been previously lost by other mechanisms;
e.g., through a very efficient stellar wind (which, on the other hand, leads to the binary expansion), 
or in an explosive merger \citep{Justham06_grb}.
Another suggested scenario, which does not have the energy-balance problem, is that a donor star was initially 
an intermediate mass star and the fossil magnetic fields provided strong magnetic braking that 
brings the binary to the onset of mass transfer \citep{Justham06_lmxb}.
It is crucial to understand that the predicted formation rates for suggested
scenarios can vary by several orders of the magnitude,  
due to the huge uncertainties in stellar evolution (winds rates, supernova kicks),
the initial binary parameters (periods, mass ratios) and in the binary evolution (e.g., 
the common envelope efficiency).

The weakness of the scenarios described above is the difficulty to reproduce 
the observationally available relation between the period and the effective temperature for the 
short-period BH X-ray binaries.
This is inevitable if the donor of an intermediate mass \citep[][]{Justham06_lmxb},
as the theoretically obtained donors, being on MT sequence, are too hot compared to the observed ones
for the same binary periods (in other words, for the same radii).
Cold temperatures at large periods are also difficult to achieve for donors of smaller masses 
and occur only if the MT starts during the Hertzsprung gap, but in the very limited range of periods.
The most difficult to explain are systems that have periods from 7 to 12 hours.
These systems are known to have a Lithium overabundance up to 200 times the solar value (consistent with
the primordial value).
Note that Li overabundance has been found not in binaries with smallest periods within non-giant donors,
but with larger periods, c.f. the proposition on the tidal locking in close binaries been responsible
for Li production in \citet{Macc05_li}.

We propose that these systems can be explained by considering pre-main sequence donors.
In addition to the observational constraints on the donor temperatures and periods, 
we take into account that by the time when a primary star explodes and creates a black hole ($6-8 \times 10^6$ yr),
its low mass companion will not reach zero-age main sequence (ZAMS). 
In fact, at the time of $10^7$ yr, only the stars more massive than 2 $M_\odot$ will come to their MS \citep{Dant94_prems}.
Therefore any binary systems with a BH (and also with a neutron star) and a low-mass MS star
will unavoidably evolve through the stage when the formed compact object is orbited by a pre-MS star.
The radius of the star at this moment is about 3 times bigger than at ZAMS -- the main contraction
has already occurred.

\begin{figure}
\plotone{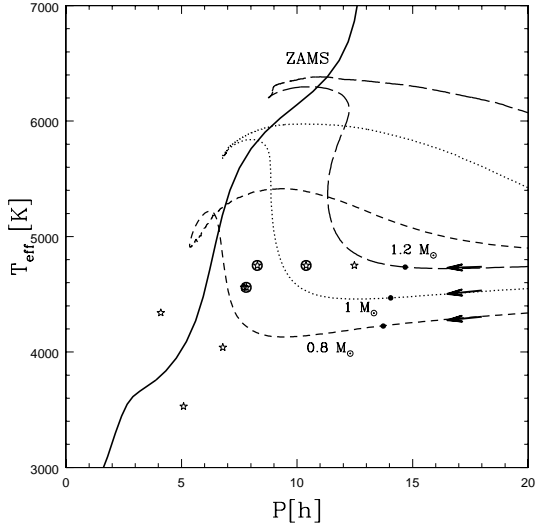}
\caption{Evolutionary tracks in the orbital period --- effective temperature plane for
donors initially of 1.2 $M_\odot$ (long-dashed line), 1.0 $M_\odot$ (dotted line) 
and 0.8 $M_\odot$ (short-dotted line) from pre-MS stage; the arrows show the direction of the
star's evolution through the plane.
Solid line shows the position of ZAMS donors.
All periods are such that stars would fill their Roche lobes if they have a 7 $M_\odot$ BH companion.
Dots mark the position of the donors at the approximate moment of BH formation, $7\times 10^6$ yr. 
``Star'' symbols are the earliest spectral classes of donors in 
observed BH X-ray binaries given in \citet{Casares05_bhm}. 
BH X-ray binaries with Li overabundance are marked with circles.
\label{pteff}
}
\end{figure}

On Fig.~\ref{pteff} we show the evolutionary tracks of 1.2, 1.0 and 0.8 $M_\odot$ stars 
in the period --- effective temperature plane, 
where the periods are such that the stars would fill their Roche lobe
with  a 7 $M_\odot$ companion.
The evolution of stars is started from their Hayashi tracks.
While stars evolve, they move from the right to the left, raising their temperature
as they approach MS. During MS, stars perform small loops on the left from ZAMS line.
Leaving  MS, stars go from left to right and have much 
higher temperatures than at the pre-MS stage.

It can be seen that, for the same periods, 
the difference between the hottest possible observed effective temperatures 
in Galactic BH X-ray binaries and those of the MS stars is very large, about 1000 K.
One can argue that a star of 0.8 $M_\odot$ passes relatively close to the 
region of interest when it becomes a giant,
however a star of this mass leaves the MS after twice the Hubble time!
After the Hubble time, only $\sim 0.95\ M_\odot$ star of solar metallicity will leaves the MS.
With a lower metallicity, Z=0.001, stars evolve faster and less massive stars can become giants
within the Hubble time. 
However, it does not help, as in this case ZAMS line is shifted upwards and, for the
same periods, stars are hotter by about 1000 K than in the case of Z=0.02.   
If a donor starts the MT during early MS life-time, then its evolutionary track during MT is going to the left and down
from its MS, the track is almost parallel to ZAMS line on Fig.~1.
This explains the observed BH X-ray binary with the shortest period, XTE J1180+480 with $P_{\rm rot}=4.1$~h.
If a star is well evolved from the MS, then it may follow the subgiant track and evolve to the right,
but it will have too high  temperature for the required period.
When the star has evolved from MS, but not too far, then its track is similar to the behavior during the MT on MS,
but it is hard to get far down to cold temperatures within the Hubble time.
We must stress that, as becomes evident from the figure, 
the orbital period of a pre-MS star at the time of SN explosion has to be at least $\sim 15$h, 
and therefore we should eliminate from the consideration the periods below $\sim 15$h for MS stars, too.

In what follows, we outline how the magnetic braking operates
in the pre-MS stars, show the evolution of binaries with a BH and a pre-MS star
through the MT stage and analyze how those systems appear.

\section{Magnetic braking at pre-main sequence stage}

General prediction of the dynamo theory 
suggests that activity should scale with the dynamo number $N_{\rm D}$ 
\citep[e.g.,][]{Parker71_dynamo,Hinata89_dynamo,Meunier97_dynamo}. 
The simple model of a dynamo operating through the convective zone 
predicts that the dynamo number is related to the Rossby number as
$N_{\rm D}\sim Ro^{-2}$ , where Rossby number 
is defined as $Ro=P_{\rm rot}/\tau_{\rm c}$ 
($\tau_{\rm c}$ is the convective turnover time and $P_{\rm rot}$ is the rotation period) 
\citep{Noyes84_dynamo}.
The corresponding correlation between activity and the Rossby number
was shown in a number of studies devoted to activity in late-type MS 
stars and has been used to interpret X-ray emission of MS \citep{Montesinos01_dynamo} and 
pre-MS stars \citep{Pizzolato03_rossby,Massi06_Rossby}, even though this correlation
has a large scatter \citep{Preibisch05_dynamo}.

\begin{figure}
\plotone{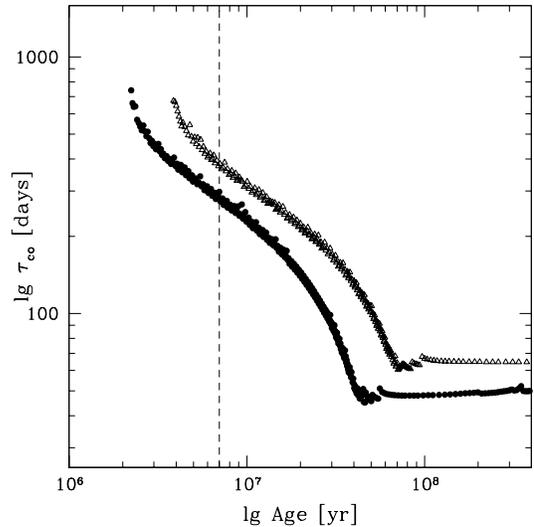}
\caption{The envelope global convective time 
$\tau_{\rm c}$ for 0.8 (open triangles) and 1 $M_\odot$ (solid circles)
stars. The dashed line shows the approximate time of the supernova explosion
of the primary. Convective times are calculated starting from the
moment when pre-main sequence stars developed a radiative core.
Flat platos correspond to the main sequence stage.
\label{tau}
}
\end{figure}

On Fig.~\ref{tau} we show the envelope convective turnover time, starting from the moment
when a contracting pre-MS star develops small radiative core, and through the MS stage:
$\tau_{\rm c}$ is decreasing as the star is contracting toward the MS and
remains almost the same as the star evolves through the MS.
This figure illustrates how the correlation between $\tau_{\rm c}$ and magnetic field strength
explains naturally high magnetic fields observed in pre-MS stars and lower magnetic fields
observed in MS stars. 
As $B\propto \sqrt N_{\rm D}$, a simple estimate shows that the field of a fast rotating young star
with the $P_{\rm rot} = 1$ d is 400 times stronger than that of the Sun with the $P_{\rm rot} = 24$ d:
$B^0/B_\odot \approx 400$. Indeed, the magnetic field on Sun is measured to be 
2~G, whereas the fields detected in T Tauri stars reach up to 5~kG \citep{2005MNRAS.358..632O,2005MNRAS.358..977S}, 
and could be higher at earlier stages. E.g., consider TW Hyd with measured magnetic field of 2.5~kG  \citep{2005ApJ...635..466Y}.
It has $T_{\rm eff}=4126\pm24$~K and luminosity $\lg L/L_\odot = -0.62\pm0.11$.
Comparing these values with evolutionary tracks, we estimate TW Hyd age as  $\sim 2\times10^7$ years.
At this age $\tau_{\rm c}$ is several times less than its maximum value.

In our study we adopt the connection between the surface magnetic field $B_{\rm s}$
and the Rossby number as:

\begin{equation}
\frac{B_{\rm s}}{B_{\rm s}^0}= f_{\rm B} \frac{\tau_c}{800^{\rm d}} \frac{1^{\rm d}}{P_{\rm rot}} ,
\end{equation}
where we take the primordial magnetic field $B_{\rm s}^0 = 10$ kG; 
when dynamo mechanism starts, $\tau_{\rm c} = 800$ d and the rotation period is 1 d; 
$f_{\rm B}$ is a scaling factor that allows us to vary $B^0$.
This ratio is similar to commonly accepted relation  ${B_{\rm s}}\propto \Omega$,
where $\Omega$ denotes the stellar rotation rate; used for
the main sequence stars \citep[e.g.,][]{Mestel87_mb}, where $Ro$ does not change strongly throughout the
life on the main sequence.

The magnetic braking, or the loss of the angular momentum from the star, is
provided through the coupling of the stellar wind with star's magnetic field. 
Young stars can lose their material at a very high rate, 
$4 \times 10^{-8}\ M_\odot$/yr  \citep[e.g.,][]{Lorenzetti06_wind}.
In weak T Tauri stars the mass loss rate is as low as $10^{-10}-10^{-11}\ M_\odot$/yr
\citep{Andre92_wind}. Overall, T Tauri stars with several kG magnetic field 
would have $10^{-8}-10^{-9}\ M_\odot$/yr wind \citep{Paatz96_wind}.
During the pre-MS stage, the wind strength decreases by a factor which is comparable to
the decrease in the magnetic field strength.
As the mass loss rate $M_{\rm w}$ and strength of the magnetic field are
correlated, even though not necessarily physically connected,
we consider their dependence in a simple parametrized form:
 
\begin{equation}
\dot M_{\rm w} = 10^{-9} M_\odot/{\rm y} f_{\rm w} \frac{B_{\rm s}}{B_{\rm s}^0} \ ,
\end{equation}
where $f_{\rm w}$ is a parameter in the range of 0.1 to 10.

For the model of the magnetic braking, we adopt the standard set of assumptions:
(i) that magnetic field is dipolar and, accordingly, drops with distance as 
$B(r)=B_{\rm s} R^3/r^3$ ($R$ is the radius of the star);
(ii) wind corotates out to the Alfv\'en radius, where the magnetic
pressure and ram pressure are balanced; (iii) wind-loss rate is the same at the surface and 
at the Alfv\'en surface (for simplicity, no wind is trapped 
within the ``dead zone'' like considered in \cite{Mestel87_mb}); 
(iv) the wind velocity at the star surface  is of order of the escape speed.
The resulting rate of change of angular momentum due to magnetic braking is
\citep[for the derivation see also, e.g.,][]{Justham06_lmxb}:

\begin{equation}
\dot J_{\rm MB} = -\Omega B_{\rm s} R^{13/4} \dot M_{\rm w}^{1/2} \left(GM\right)^{-1/4} ,
\end{equation}
where $M$ and $R$ are the mass and the radius of the pre-MS star.

\section{Binary evolution}

We used the binary evolutionary code with implemented implicit treatment
of the mass transfer \citep{Ivanova04_mt}. 
We adopted the magnetic braking described
above and considered binaries with a 7 $M_\odot$ BH with 1.0 and 0.8 $M_\odot$ pre-MS companions.
We started the evolution with several values of the period at the onset of MT, 9.5 h 
and 13 h for a binary with 1$M_\odot$ companion and 13.3 h with 0.8 $M_\odot$ companion.
For each case, we varied the initial magnetic fields ($f_{\rm B} = 1,2,5,10$ in the eq.~(1)). In total,
we calculated 12 evolutionary tracks.

\begin{figure}
\plotone{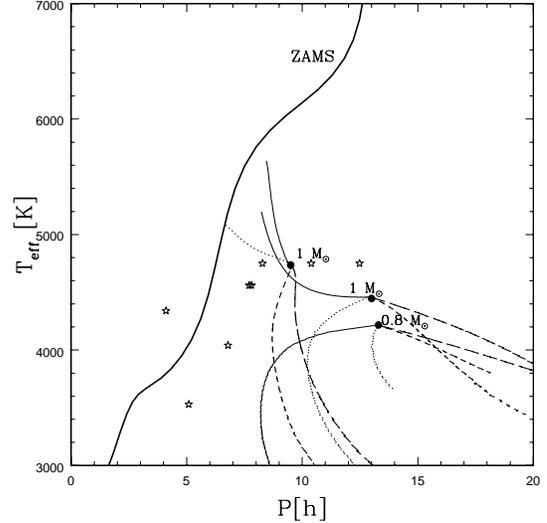}
\caption{Evolutionary tracks in the orbital period --- effective temperature plane for donors
of 0.8 $M_\odot$ and 1 $M_\odot$  (two case with different initial periods are considered in the latter case).
All initial binaries are evolved with 4 values of magnetic pole strength ($f_{\rm B} = 1, 2, 5$ and 10,
corresponding lines are solid, dotted, short-dashed and long-dashed),
the resulting evolution tracks rotate counter-clockwise around the initial binary position (marked with large solid dots).
Other notations are as on Fig.~\ref{pteff}.
\label{fig-mt}
}
\end{figure}

The results are shown on Fig~\ref{fig-mt}.
The end of the tracks correspond to either the stop of the mass transfer when the donor becomes detached; or when
the donor reaches ZAMS. There are several reasons for the donor to become detached - it can contract
when it approaches the MS (in the case of low magnetic field), or magnetic braking stops to operate when
the star becomes fully convective: we assume the dipole field generation stops to be effective in this case.
In the latter case, the system is unlikely to reappear in the future as an X-ray system since the period is too wide
for gravitational waves to shrink it, and the magnetic braking is switched off.
The tracks shown demonstrate that it is possible to keep the binary on the onset of MT
during the pre-MS contraction of the donor and that systems appear in the same region
as the observed short-period BH X-ray binaries.

In particular, the evolution of binaries with a low-field pre-MS stars
is not much different from the evolutionary path they would have
if they would be just barely at the contact and not losing much of their mass; 
they evolve to shorter periods and higher temperatures.
As the pole-strength increases, the evolutionary path of the star makes counterclockwise rotation
in the $P$--$T_{\rm eff}$ plane, and for extremely high fields,
a star is so much out of the thermal equilibrium 
that starts not only to cool down as it loses its mass, but also expands.

We have estimated roughly the maximal post-explosion periods for which a binary will start the MT
during pre-MS stage.
As expected, the higher initial $B_{\rm S}^0$, the wider initial binary can be at the moment 
of the SN explosion and still start the MT during the pre-MS stage. 
For example, if $f_{\rm B} = 10$, 
$P_{\rm SN} \approx 1.8$~d in a binary with 7 $M_\odot$ BH and 1 $M_\odot$ pre-MS star, 
and if $f_{\rm B} = 3$, then $P_{\rm SN}$ is only 1.1~d.
We note that those periods are about the same or higher than those of a BH / MS star binary
that starts MT during the MS.

\section{Discussion}

We may list several possible criticisms of the proposed scenario:
First, the interval during which such binaries can exist after the star formation is small,
about several $10^7$ yr at best, and, accordingly, must be associated 
with a relatively young stellar population. Such a trend is not reported.
On the other hand, if the supernova kick was as small as 50 km/s, a binary
can be several hundreds parsecs  away from the place of it's birth.
Short life-time however suggest that such systems can not be present among
observed low mass X-ray binaries in elliptical galaxies and the bulges of some spiral galaxies,
where there is no star formation  \citep{2004PThPS.155...49G}.
Second, observed short-period LMXBs with BH donors are transient.
Except for the systems with a low magnetic field, all other our systems 
have relatively high MT rates and should be seen as persistent, unless
the criterion for the transiency would differ significantly in the case
of unusually strong magnetic fields.
And last, the life-time of the system during X-ray phase is shorter than it would be
if a donor is at MS, about several $10^7$~years versus several  $10^8$ or even $10^9$~years.
The short life-time will affect the prediction on how many binaries can be observed.
We do not attempt to estimate the formation rates properly, due to the huge uncertainties 
that affect the rates for standard scenarios at first place. But we want to note that 
neither supernova kicks or the energy balance during the common envelope  
would reduce the formation rates compared to the standard formation channel
with low-mass companions, as they act in the same way. 
On the other hand, the large range of post-explosion binary periods
may increase the formation rate. The resulting detection rates might be similar
to those predicted for a standard channel with the same set of assumption for kicks
and common envelope efficiency.

In the conclusion we want to emphasize again that
systems with a BH or a NS accretor and a pre-MS star will exist, it is not possible
to avoid their formation.
Being affected by strong magnetic field, those binary systems will shrink 
faster than during the MS evolution and this will alter the 
period distribution of the post-supernova systems.
Short period binary systems with a compact object and a low mass companion
are very likely to start the MT during the pre-MS stage of the secondary.
In fact, if the post-explosion period is about a day, the system will definitely start the MT
during the pre-MS stage even with a moderate magnetic field of the donor.
If the post-explosion period is larger, then the system is unlikely to start the MT during both pre-MS and MS stages.
As one of the consequences, when BH-MS systems are considered, the distribution of binary periods must take into account
that pre-MS stars were significantly bigger at the time of the supernova explosion.
This is true as well for NS-MS systems and should be taken into account when, e.g., the formation
of binary millisecond pulsars is studied.

During the MT, the examined BH-pre-MS systems have the effective temperatures and periods
similar to the observed BH X-ray binaries systems.
As little or none nuclear burning occurs in donors at these ages, they will have
high (and up to primordial) Li abundance, similar to detected in donors of BH X-ray binaries.
It is interesting to mention that the donor where some CNO-processing was seen, XTE J1118+480 \citep{Hasw02_bh},
has the period of 4.1~h. It is therefore located on the left of the ZAMS line (Fig.~\ref{pteff}) 
and is explained with the evolution of a MS donor at the start of the MT.
In addition to 3 BH X-ray binaries with Li excess described earlier, there are two others.
One is a neutron star system Cen X-4 \citep{1994ApJ...435..791M}.
With the orbital period of 15.1 hours, it still can be explained by our model.
We however expect that fewer NS binaries can be affected by the proposed mechanism, as 
the formation of a NS occurs later than that of a BH. Another object is a BH X-ray binary 
V404 Cygni \citep{1992Natur.358..129M}.
With the orbital periods of 6.5 days and a giant companion, this binary 
is well beyond pre-MS stage and Li excess here must have another origin.
E.g., in low-mass giants Li can be created via Cameron-Fowler mechanism
\citep{1999ApJ...510..217S} with the possible enhancement 
in the case of tidally locked binaries \citep{2006ApJ...641.1087D}

More detailed investigation of magnetic fields in pre-MS stars and their dependence on the stars age and rotation 
may help greatly for the improving of the described above simple magnetic braking model and
for the corresponding understanding of LMXBs formation and evolution.
We also predict that observations of X-ray activities of BH X-ray binaries during the quiescence may confirm
or disproof the hypotheses on pre-MS nature of their donors.

\section*{Acknowledgments}

Author acknowledges helpful discussion with S.~Justham.


\end{document}